\documentclass[aps,prb,twocolumn,superscriptaddress,showpacs]{revtex4}
\usepackage{graphicx}
\usepackage{amsmath}
\usepackage{epsfig}

\newcommand{\beq}{\begin{eqnarray}}
\newcommand{\eeq}{\end{eqnarray}}

\begin{document}

\title{Surface Plasmons in a Metal Nanowire Coupled to Colloidal Quantum Dots: Scattering Properties and Quantum Entanglement}
\author{Guang-Yin Chen}
\affiliation{Department of Physics and National Center for
Theoretical Sciences, National Cheng-Kung University, Tainan 701,
Taiwan}
\affiliation{Advanced Science Institute, RIKEN, Wako-shi,
Saitama 351-0198, Japan}
\author{Neill Lambert}
\affiliation{Advanced Science Institute,
RIKEN, Wako-shi, Saitama 351-0198, Japan}
\author{Chung-Hsien Chou}
\affiliation{Department of Physics and National Center for
Theoretical Sciences, National Cheng-Kung University, Tainan 701,
Taiwan}
\author{Yueh-Nan Chen}
\email{yuehnan@mail.ncku.edu.tw} \affiliation{Department of Physics
and National Center for Theoretical Sciences, National Cheng-Kung
University, Tainan 701, Taiwan}
\author{Franco Nori}
\affiliation{Advanced Science Institute, RIKEN, Wako-shi, Saitama
351-0198, Japan}
\affiliation{Physics Department, University of
Michigan, Ann Arbor, MI 48109-1040, USA}
\date{\today}

\begin{abstract}
We investigate coherent single surface-plasmon transport in a metal
nanowire strongly coupled to two colloidal quantum dots. Analytical
expressions are obtained for the transmission and reflection
coefficients by solving the corresponding eigenvalue equation.
Remote entanglement of the wave functions of the two quantum dots
can be created if the inter-dot distance is equal to a multiple
half-wavelength of the surface plasmon. Furthermore, by applying
classical laser pulses to the quantum dots, the entangled states can
be stored in metastable states which are decoupled from the surface
plasmons.
\end{abstract}

\maketitle

%%%%%%  VERSION 16 JUNE 2006 - as submitted
% -------------------------------------------------------------
% Use the \preprint command to place your local institutional report
% number in the upper righthand corner of the title page in preprint mode.
% Multiple \preprint commands are allowed.
% Use the 'preprintnumbers' class option to override journal defaults
% to display numbers if necessary
%\preprint{}

% repeat the \author .. \affiliation  etc. as needed
% \email, \thanks, \homepage, \altaffiliation all apply to the current
% author. Explanatory text should go in the []'s, actual e-mail
% address or url should go in the {}'s for \email and \homepage.
% Please use the appropriate macro foreach each type of information

% \affiliation command applies to all authors since the last
% \affiliation command. The \affiliation command should follow the
% other information
% \affiliation can be followed by \email, \homepage, \thanks as well.

%\preprint{APR 2004-XX}

\section{INTRODUCTION}

Colloidal quantum dots (QDs) are fluorescent core-shell
semiconductor nanocrystals with tunable luminescence properties
(e.g., broad excitation spectra, narrow emission spectra, and
size-dependent emission), which have recently attracted much
attention for their ability to act as photon
detectors\cite{konstantatos}.

When a light wave strikes a metal surface, it can excite a
surface-plasmon-polariton: a surface electromagnetic wave coupled to
plasma oscillations. Recently, the concept of plasmonics, in analogy
to photonics, has received much attention since surface plasmons
(SPs) reveal strong analogies to light propagation in conventional
dielectric components\cite{Zia,Bliokh,Savel'ev}. For example, it is
now possible to confine them to subwavelength
dimensions\cite{Barnes} leading to novel approaches for waveguiding
below the diffraction limit\cite{Gramotnev}. The combination of
subwavelength confinement, single-mode operation, and the relatively
low-power propagation loss of SP polaritons could be used to
miniaturize existing photonic circuits\cite{Bozhevolnyi}.
Furthermore, the strong coupling between SP and emitters\cite{gomez}
can be utilized to enhance infrared photodetectors\cite{Chang}, the
fluorescence of QDs\cite{Hwang}, and light transmission through
metal nanoarrays\cite{Gao}. High-field surface plasmon confinement
was also used to demonstrate an all-optical
modulator\cite{Pacifici}, to provide an extra degree of freedom for
information storage\cite{Zijlstra}, and to estimate the reflectivity
of structures or surface roughness\cite{Dragnea}.

In a related context, advances in quantum information science have
promoted an experimental drive for physical realizations of highly-entangled states\cite{Stolze}. Successes have been obtained within
quantum-optical and atomic systems\cite{Stolze}. However, due to
scalability requirements, solid-state realizations of such phenomena
are promising\cite{Costa,Oliver,gywat}. Furthermore, while initial
attention has been focused on the coupling between nearby qubits
with local interactions\cite{Berkley,Steffen,You}, entangling
arbitrary pairs of remote qubits is still an important goal. Circuit
quantum electrodynamics (QED), for example, is one promising
candidate to couple two distant qubits via a cavity
bus\cite{Majer,You}.

Motivated by these recent developments in plasmonics and quantum
information science, we propose a novel scheme that can entangle
two remote QDs coupled to a metal nanowire. The idea is inspired
by recent experiments showing single surface plasmons in metallic
nanowires coupled to QDs\cite{Akimov}. We use a
real-space Hamiltonian\cite{JT} to treat the coherent
surface-plasmon transport in the wire coupled to two dots.
The transmission and reflection spectra of SPs  for both
single QD and double QDs cases in the \textit{non-linear quadratic
dispersion regime} has been studied in our previous
work\cite{wei}.  In that work the scattering spectra reveals a
Fano resonance due to the interference between localized and
delocalized SP channels. Here, we focus on the entanglement
generation through the scattering of SPs in \textit{linear
dispersion relation regime}. In the limit of infinite
Purcell factor, $P\rightarrow\infty$, we find that maximally
entangled states can be created if the separation between the two
dots is equal to an integer multiple half-wavelength of the
optical plasmon. Furthermore we show the entangled state
information can also be stored in metastable states, which are
decoupled from the surface plasmons, by separately applying
classical laser pulses to each QD. The storage efficiency of the
entangled states is equal to ($1-1/P$).

A similar scheme was also recently proposed independently in
Ref.~[\onlinecite{spain1,spain2}], who
 solved a similar setup using a master equation
description of the two quantum dots.  Their findings are consistent
with our results, with the maximum of the concurrence occurring at
the same values of the dot distance.   Furthermore, they show the
concurrence behavior as a function of time, for a single initial
excitation, which illustrates that indeed the concurrence exists on
an experimentally accessible time-scale. The approaches used are
very different: we use a real-space scattering formulation, in
contrast with their master equation. We stress that the reflection
and transmission coefficients computed here can be directly accessed
experimentally, which is not easy to do by other methods.
Furthermore, we have considered the important effects of Ohmic loss.
Moreover, we propose a way to store the entanglement produced in
this manner.
\section{PLASMON SCATTERED BY TWO QUANTUM DOTS}
\begin{figure}[th]
\includegraphics[width=\columnwidth]{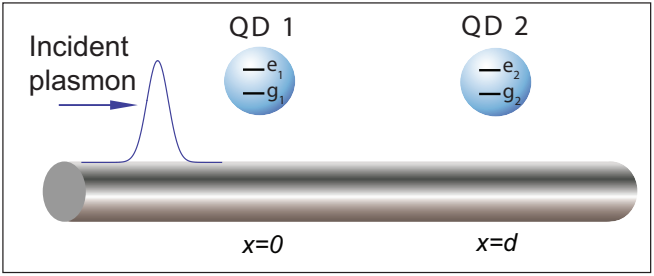}
\caption{(Color online) Schematic diagram of a metal nano-wire coupled to two QDs.
A single surface plasmon injected from the left is coherently
scattered by the dots. }
\end{figure}
When a colloidal semiconductor QD is placed close to a metal
nanowire, a strong coupling between the QD exciton and SP can
occur\cite{Chang1, gy, Dzsotjan}, as in traditional cavity QED\cite{Khitrova}.
As shown in Fig.~1, the system we consider here is composed of two
nominally identical QDs, both with energy spacing $\hbar\omega_{0}$,
separated a distance $d$, near a cylindrical metal nanowire.
 We assume that a surface plasmon emitted from a source QD is incident from the left with energy
 $E_{k}=v_{\textrm{g}}k$. Here, $v_{\textrm{g}}$ and $k$ are the group velocity and wavevector of the incident
 SP, respectively.  Since the SP is propagating on the surface of the metal
nanowire, it inevitably suffers from dissipation (such as Ohmic
loss). As a result, the source of the incoming plasmon (e.g. an
additional QD) must be placed close to the first QD to minimize
the initial losses (i.e. those which occur before the plasmon
reaches the two-dot region). We also propose a practical way to
further minimize all the dissipation affects in the next section.
The dissipation terms are included as an imaginary energy in the
non-Hermitian effective Hamiltonian~\cite{chang2,Shen}, which can
be written as
\begin{eqnarray}
&&H=\sum_{j=1,2}\hbar \left[ \omega _{0}-i\left(\frac{\Gamma'}{2%
}\right)\right]\sigma _{e_{j},e_{j}}+\int dk~\hbar v_{\textrm{g}}|k|a_{k}^{\dag }a_{k}\notag\\
&&-\hbar g\int
dk~\left[\left(\sigma^{(1)}_++\sigma^{(2)}_+e^{ikd}\right)a_{k}+\textrm{h.c.}\right]\notag\\
 &&-i\hbar
\frac{\sin(k_{0}d)}{2k_{0}d}\gamma _{0}\left(\sigma^{(1)}_+\sigma^{(2)}_-+\sigma^{(1)}_-\sigma^{(2)}_+\right),  \notag \\
\label{H1}
\end{eqnarray}
where $\sigma _{e_{j},e_{j}}=|e_{j}\rangle \langle e_{j}|$
represents the diagonal
 element and $\sigma^{(j)}_+=|e_{j}\rangle \langle g_{j}|$ represents the off-diagonal
of the $j$-th QD operator, and $a_{k}^{\dag }$ is the creation
operator of the surface plasmon. Here,
$k_0=\omega_{0}/v_\textrm{g}$, and $g$ is the coupling constant
between the excitons and SP. The term in the last line,
\begin{equation}
\Gamma^{\textrm{SR}}=\bigg[\frac{i\sin(k_{0}d)}{2(k_0d)}\bigg]\gamma_{0},
\end{equation}
is the contribution from the collective decay (Super-Radiance,
hence the super-index ``\textrm{SR}") effect\cite{ynchen}, with
$\gamma _{0}$ being the exciton decay rate into free space, and
\begin{equation}
\Gamma'\equiv\gamma_0+\Gamma_{0},
\end{equation}
is the total dissipation including the decay rate into free space
$\gamma_0$ and other dissipative channels (for example, the Ohmic
loss) $\Gamma_0$.  We can include the Ohmic loss of plasmons
during the scattering process in this way because a loss of the
plasmon is equivalent to a lose of the excitation in the quantum
dots.  In addition,
\begin{equation}
\delta\equiv \frac{E_k}{\hbar}-\omega_{0},
\end{equation}
is the detuning between the incident SP energy with $E_k$ and the
QD exciton energy $\omega_{0}$.

The validity of this non-Hermitian form relies on the fact that
$\Gamma'$ is small and that we only consider one excitation. Hence
we do not need to consider the effect of the quantum jump terms
one usually needs for a full description of the
system\cite{chang2, Meystre}. However, this implies that there is
a certain time scale of decay of the highly-entangled state into
other modes, even if they are not emitted into the
surface-plasmonic environment (though typically the Purcell factor
is large, so emission into SPs will dominate any decay process).
This reinforces the need to transfer the fragile entangled state
to a metastable state.

Since we are only interested in the case where the incident SP is
nearly-resonant with the two QDs, we can rewrite $\int dk~\hbar
v_{\textrm{g}}|k|a_{k}^{\dag }a_{k}$ as $\int dk~\hbar
v_{\textrm{g}}k(a_{R,k}^{\dag }a_{R,k}+a_{L,k}^{\dag }a_{L,k})$
and $(\sigma^{(1)}_++\sigma^{(2)}_+e^{ikd})a_{k}$ as
$(\sigma^{(1)}_++\sigma^{(2)}_+e^{ikd})(a_{R,k}+a_{L,k})$.
Transforming Eq.~(\ref{H1}) into real space, one obtains
\begin{eqnarray}
&&H=\hbar \int dx\left\{-iv_{\textrm{g}}c_{R}^{\dag }(x)\frac{\partial }{\partial x}%
c_{R}(x)+iv_{\textrm{g}}c_{L}^{\dag }(x)\frac{\partial }{\partial
x}c_{L}(x)\right. \notag
\\
&&\left. +\hbar g\sum_{j=1,2}\delta (x-(j-1)d)\left[c_{R}^{\dag
}(x)\sigma^{(j)}_-+c_{R}(x)\sigma^{(j)}_+\right.\right.  \notag \\
&&\left.\left.+c_{L}^{\dag
}(x)\sigma^{(j)}_-+c_{L}(x)\sigma^{(j)}_+\right]\right\}
+\sum_{j=1,2}\hbar\left[  \omega _{0}-i
\left(\frac{\Gamma'}{2}\right)\right]\sigma
_{e_{j},e_{j}} \notag \\
&&-i\hbar \frac{\sin(k_{0}d)}{2k_{0}d}\gamma
_{0}\left(\sigma^{(1)}_+\sigma^{(2)}_-+\sigma^{(1)}_-\sigma^{(2)}_+\right),
\label{H2}
\end{eqnarray}
where $c_{R}^{\dag }(x)$ [$c_{L}^{\dag }(x)$] is a bosonic
operator creating a right-going (left-going) surface plasmon at $x
$.

The eigenstate with energy matching the incoming plasmon
$E_{k}=v_{\textrm{g}}k$ can be written as
\begin{eqnarray}
|E_{k}\rangle  &=&\int dx\left[\phi _{k,R}^{+ }(x)c_{R}^{\dag
}(x)+\phi _{k,L}^{+}(x)c_{L}^{\dag }(x)\right]|g_{1},g_{2}\rangle
\left| 0\right\rangle
_{\textrm{sp}}  \notag \\
&&+\sum_{j=1,2}\xi_{k_{j}}\sigma^{(j)}_+|g_{1},g_{2}\rangle \left|
0\right\rangle _{\textrm{sp}}, \label{ek}
\end{eqnarray}
where $\xi_{k_{j}}$ is the probability amplitude that the
$j$-th QD absorbs the surface plasmon and jumps to its excited
state, and $|g_{1},g_{2}\rangle \left| 0\right\rangle
_{\textrm{sp}}$ means that both, QD-1 and QD-2, are in their ground
states with no SPs. %Here $\sigma^{(j)}_+$ denotes the raising
%operator, which raises the $j$-th QD from the ground to the excited
%state, and $c_{R}^{\dag }(x)$
%[$c_{L}^{\dag }(x)$] is a bosonic operator creating a right-going
%(left-going) surface plasmon at $x $.
For a single SP incident from the left, the scattering amplitudes
$\phi _{k,R}^{+}(x)$ and $\phi _{k,L}^{+}(x)$ take the form
\begin{equation}
\left\{
\begin{array}{l}
\phi _{k,R}^{+ }(x)\equiv \exp(ikx)\left[\theta (-x)+a~\theta
(x)\theta
(d-x)+t~\theta (x-d)\right], \\
\phi _{k,L}^{+ }(x)\equiv \exp(-ikx)\left[r~\theta (-x)+b~\theta
(x)\theta (d-x)\right],
\end{array}
\right.
\end{equation}
where $\theta(x)$ is the unit step function which equals unity when
$x\geq0$ and zero when $x<0$. Moreover, $t$ and $r$ are the
transmission and reflection amplitudes, respectively, and
$\left[a~\exp(ikx)\theta (x)\theta (d-x)\right]$ and
$\left[b~\exp(-ikx)\theta
(x)\theta (d-x)\right]$ represent the wavefunction of the SP between $0$ and $d$%
. By solving the eigenvalue equation $H|E_{k}\rangle
=E_{k}|E_{k}\rangle $, one can analytically obtain the reflection
and transmission coefficients of single surface plasmon
scattering~\cite{chang2,Shen}. Doing so we obtain the following
relations for the coefficients
\begin{equation}
\left\{
\begin{array}{l}
g\left(2ae^{ikd}+2be^{-ikd}\right)-\Gamma^{\textrm{SR}}~\xi_{k_{1}}=\left(\delta+\frac{i\Gamma'}{2}\right)\xi_{k_{2}}, \\
g\left(1+a+r+b\right)-\Gamma^{\textrm{SR}}~\xi_{k_{2}}=\left(\delta+\frac{i\Gamma'}{2}\right)\xi_{k_{1}}, \\
a=1+\frac{g\xi_{k_{1}}}{iv_{\textrm{g}}},~b=\frac{g\xi_{k_{2}}}{iv_\textrm{g}}\exp(ikd), \\
t=1+\frac{g}{iv_\textrm{g}}\left[\xi_{k_{1}}+\xi_{k_{2}}\exp(-ikd)\right],\\
r=\frac{g}{iv_\textrm{g}}\left[\xi_{k_{1}}+\xi_{k_{2}}\exp(ikd)\right],
\end{array}
\right.\label{coeff}
\end{equation}
The transmission and reflection amplitudes can then be determined
algebraically. In the following discussion, we refer to the decay
rate into surface plasmon modes as $\Gamma_{\textrm{pl}} = 4\pi
g^2/v_\textrm{g}$.  This is convenient way to compare the strength
of the plasmon-dot coupling and other decay channels.

\subsection{Reflection and transmission of surface plasmons}

First of all we analyze the reflection and transmission
properties.  Figure 2(a) numerically displays the transmission
coefficient $T=\left| t\right| ^{2}$ (dashed lines) and reflection
coefficient $R=\left| r\right| ^{2}$ (solid lines) for different
values of the inter-dot distance $d$. The peak positions of the
reflection coefficients sometimes deviate from the center ($
\delta =0$). Figure 2(c) shows the peak position as a function of
$kd$. The continuous green (dotted blue) curve represents the
result with (without) the super-radiant effect. As can be seen in
Fig.~2(c), not only the interference from the inter-dot
separation, but also the super-radiance affects the positions of
the peaks. This is the only real contribution the super-radiance
term makes to our results.  Its effect on the entanglement,
discussed in the next section, diminishes quickly as a function of
$d$, thus we omit it completely in that treatment (see methods).

Figure 2(b) shows that the amplitude $r$ of the reflection
coefficient $R$ is suppressed when increasing the non-radiative
loss $\Gamma _{0}$. The inset in Fig.~2(b) is the scattering
spectrum versus detuning $\delta$ of a single SP incident on two
QDs with $kd$ equal to an integer multiple of $\pi$. When
$kd=2n\pi $ [$(2n+1)\pi$], with $n$ being an integer, the phase
difference of the SP between the two QDs is in-phase ($\pi$ phase
difference). This makes the two QDs collectively act like a single
QD and results in an identical spectrum to the single QD
case\cite{chang2}. Note that the reflection coefficient
$R(\delta)$ has one minimum when $\delta <0$. Without the
super-radiant effect, the positions of the minima, $\delta _{\min
}$, can be deduced from Eq.~(\ref{coeff}) and satisfy the
following relation:
\begin{equation}
-\tan ^{2}(kd)=-4\left(\frac{\delta _{\min }}{\Gamma
_{\textrm{pl}}}\right)^{2}-\left(\frac{\Gamma'}{\Gamma
_{\textrm{pl}}}\right)^{2},\label{tanthing}
\end{equation}%
where $\Gamma_{\textrm{pl}}$ is the QD exciton decay rate into SP
modes, and $R(\delta_{\min})=0$ . If there is no reflection
($r=0$), one can say that Eq.~(\ref{tanthing}) is the resonant
tunneling condition for a photon traveling through two QDs. This
$r=0$ minimum can be seen close to $\delta=0$ in Fig.~2(b). Note
that our $R(\delta)$ and $T(\delta)$ are not symmetric around
$\delta=0$. This is because: (i) the incoming wave meets two
scatterers, instead of one; and (ii) because we are considering
the effects of super-radiance. Without the conditions (i) and
(ii), (e.g., when $d=0$) we would recover the Breit-Wigner
scattering resulting in a symmetric Lorentzian, as opposed to the
asymmetric forms we obtain for $R(\delta)$ and $T(\delta)$.

\begin{figure}[htp]
\includegraphics[width=\columnwidth]{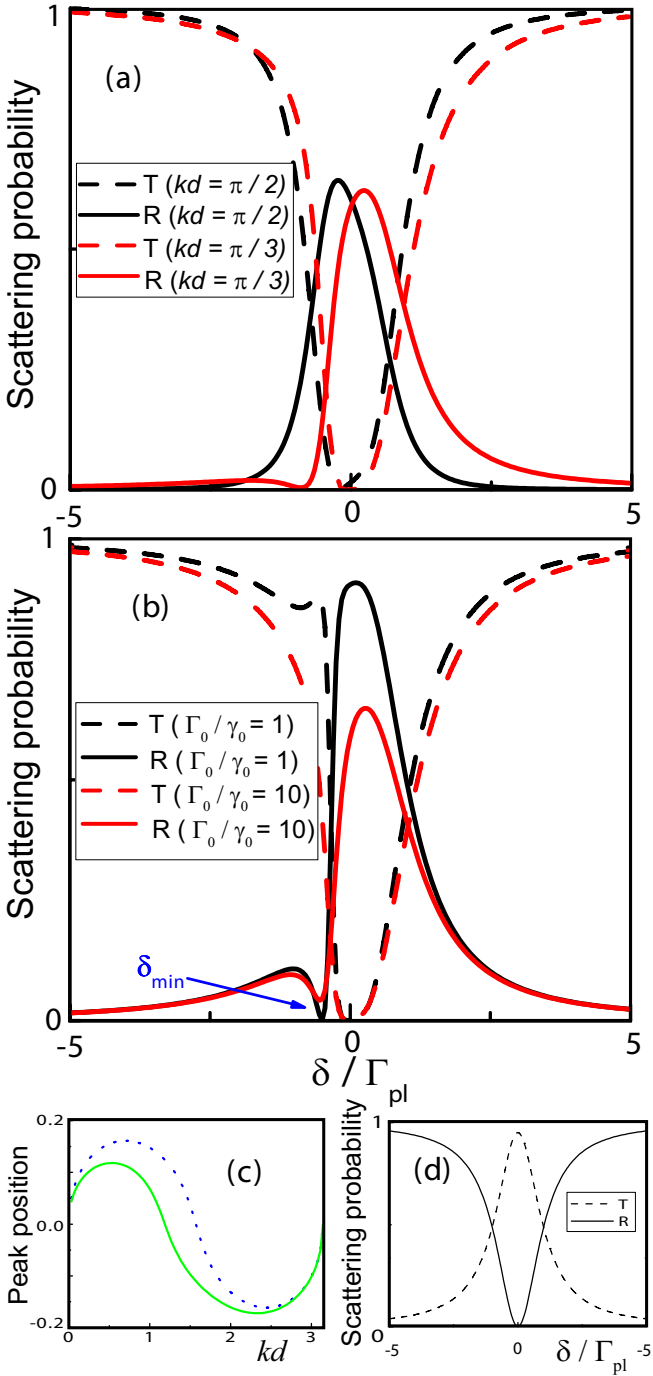}
\caption{(Color online) Transmission probabilities $T=\left| t\right| ^{2}$
(dashed lines) and reflection probabilities $R=\left| r\right|
^{2}$(solid lines) for a single surface plasmon scattered by two
quantum dots, as a function of detuning
$\protect\delta=E_k/\hbar-\omega_0 $. In the figures, $\gamma
_{0}$ and $\Gamma _{0}$ are normalized to the decay rate into the
surface-plasmon modes $\Gamma_{\textrm{pl}}$, and we have chosen
$\protect\gamma _{0}=\Gamma _{0}=0.025\Gamma _{\textrm{pl}}$ in
(a), and $kd=\protect \pi /4$, $\gamma _{0}=0.025\Gamma _{\textrm{pl}}$ in (b). (c) shows the peak positions
of the reflection probabilities as a function of $kd$. The
continuous green (dotted blue, top) curve represents the result
with (without) the super-radiant effect. (d) refers to a surface
plasmon incident on a single QD\cite{chang2}, which is also the
case for two QDs with $kd$ equal to a multiple of $\pi$. }
\end{figure}

\begin{figure}[th]
\includegraphics[width=9cm]{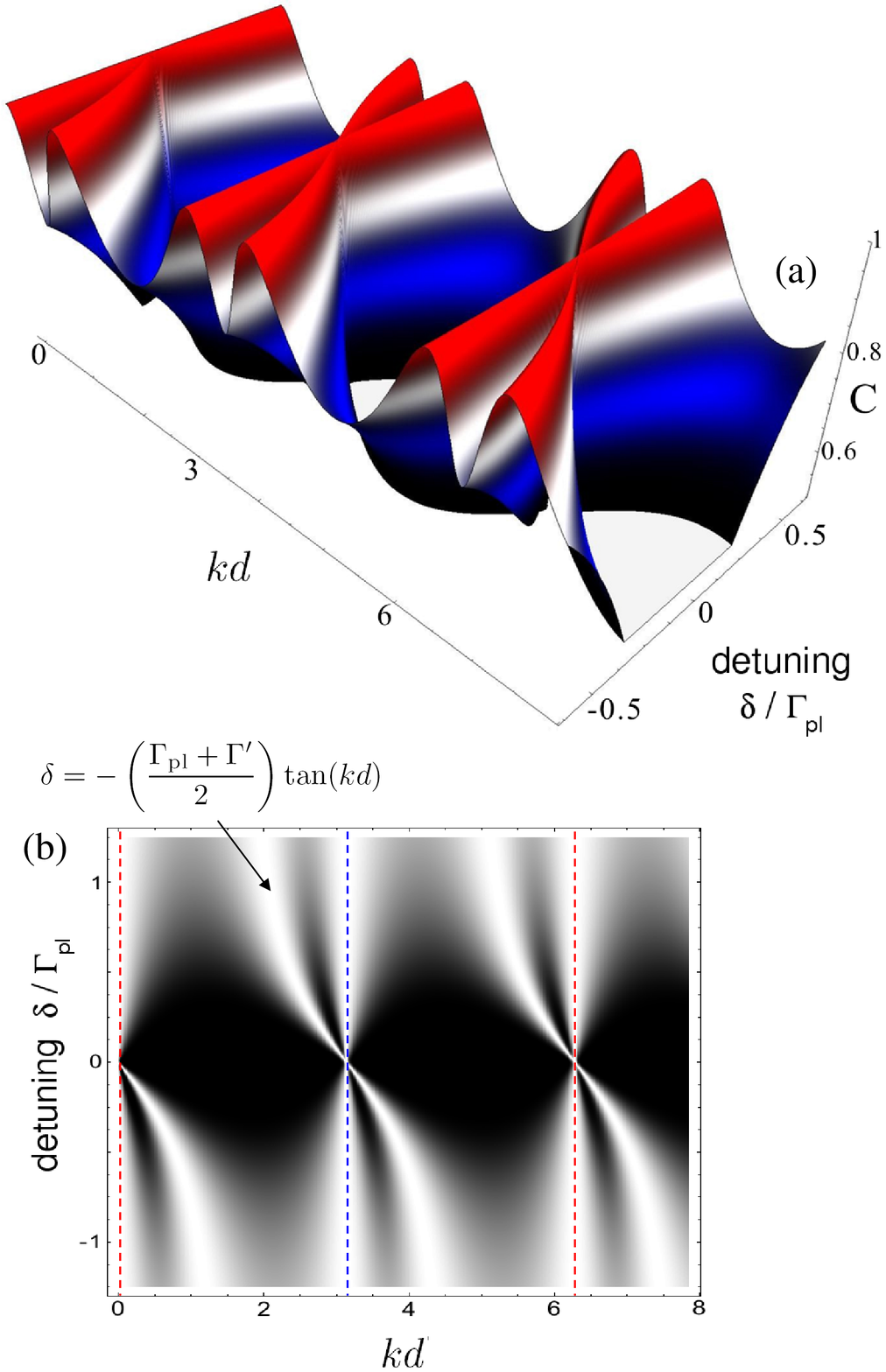}
\caption{(Color online) (a) Concurrence $C$ of the two-dot qubits as functions of
the inter-dot distance $d$ and detuning $\protect\delta $ in the limit $P\rightarrow\infty$. (b) shows
the density plot of (a): the white regions correspond to high
entanglement, with concurrence around $1$. The red vertical dashed
lines refer to $kd=2n\pi$, while the blue vertical dashed line
denotes the $kd=(2n+1)\pi$ case. Both of these conditions achieve
high entanglement.}
\end{figure}

\begin{figure}[th]
\includegraphics[width=6cm]{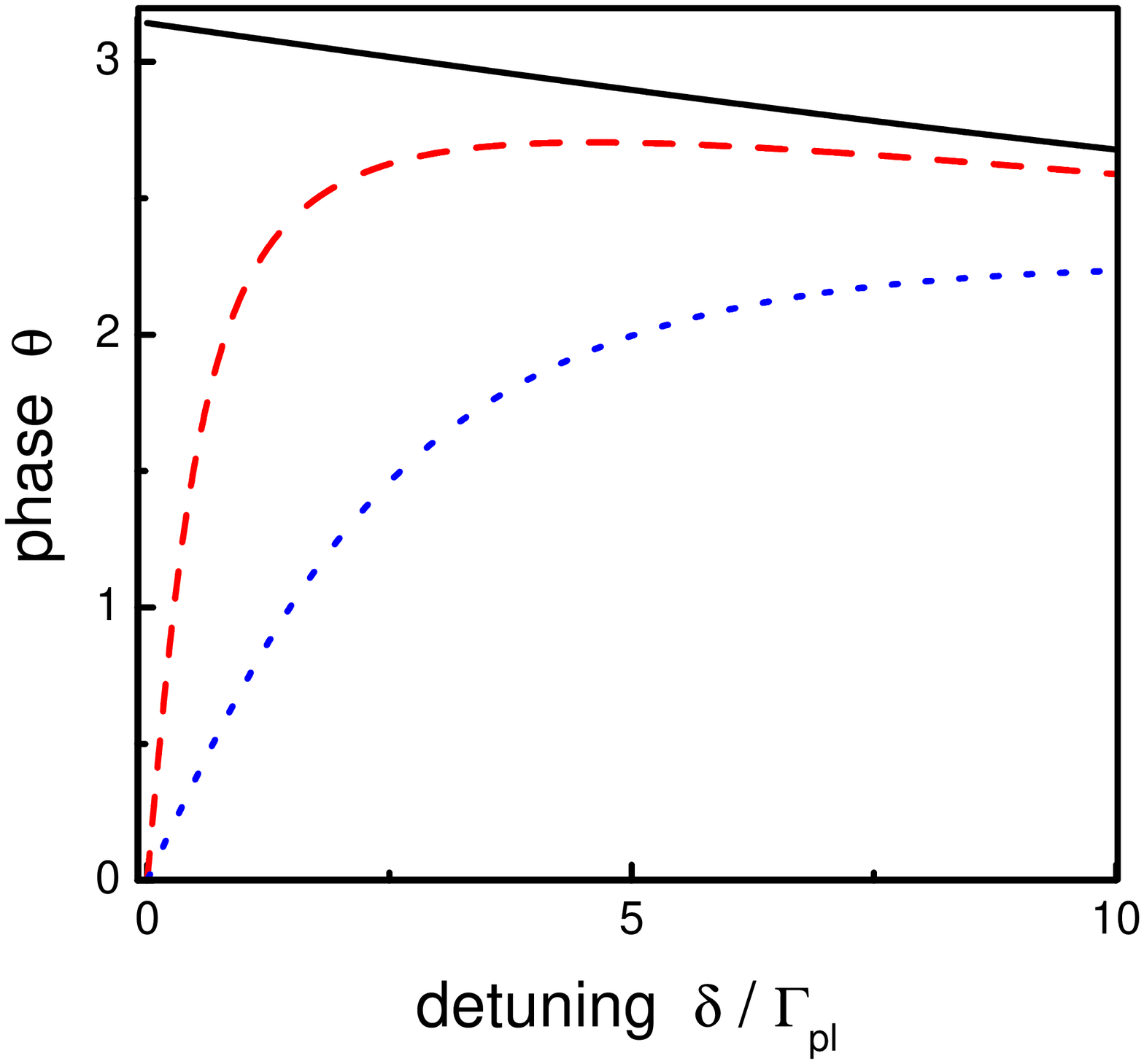}
\caption{(Color online) The phase factor $\protect\theta $ of the entangled state
$\xi_{k_{1}}|e_{1},g_{2} \rangle +e^{i\protect\theta }\cdot
\xi_{k_{2}}|g_{1},e_{2}\rangle $ in the limit of large $d$, when
super-radiance is negligible. The continuous black (top), dashed
red, and dotted blue (bottom) curves represent the results for
$\Gamma' =0,$ $0.025\Gamma _{\textrm{pl}},$ and $0.125\Gamma
_{\textrm{pl}}$, respectively.}
\end{figure}

\subsection{Entanglement generation}

The transmitted or reflected SP propagating on the surface of the nanowire can be detected by the detectors at both ends of the nanowire\cite{Akimov}.
Equations (1) and (2) oppositely imply that if there is no transmission or reflection SP detected at the two ends of the wire, the wavefunction is projected onto the state
\begin{equation}
\sum_{j=1,2}\xi_{k_{j}}\sigma^{(j)}_+|g_{1},g_{2}\rangle
\left| 0\right\rangle _{\textrm{sp}}.
\end{equation}
This means that it is possible to create entanglement between the
two dots.  However, this projection, or post-selection,
requires minimal Ohmic loss.  This can be mitigated using the
meta-stable states and waveguides discussed in the next section.
Let us first consider the limit of $P\rightarrow\infty$
($P\equiv\Gamma_\textrm{pl}/\Gamma'$). Two special cases for
achieving high entanglement are $kd=2n\pi $ and $(2n+1)\pi $,
for which the amplitude $\xi_{k_1}$ is equal to $\xi_{k_2}$ or
$-\xi_{k_2}$, respectively. In this case, the two-dot qubits
become triplet- or singlet-entangled if no surface plasmons are
detected at the two ends of the wire.

To demonstrate the degree of entanglement, Fig.~3 shows the
concurrence\cite{wootters} $C$ of the two-dot qubits as functions of
the inter-dot distance and detuning $\delta$. The concurrence
quantifies the degree of entanglement of two qubits. For our system, the density matrix of the two-dot state is a pure state density matrix, and the concurrence simply takes the form,
\begin{equation}
C=\frac{2|\xi_{k_1}|\cdot|\xi_{k_2}|}{|\xi_{k_1}|^2+|\xi_{k_2}|^2}.
\end{equation}
The red dashed line in Fig.~3(b) refers to $kd=2n\pi $, while the
blue dashed line refers to $kd=(2n+1)\pi $. In addition to the
special cases mentioned above, there are several curved
``tangent-shaped" regions satisfying the condition of high
entanglement, $C\approx1$. In the limit of large $d$, we can
neglect the effect of super-radiance and find that the equation of
these curved regions are given by
\begin{equation}
\delta =-\left(\frac{\Gamma_{\textrm{pl}}+\Gamma'}{2}\right)\tan
(kd)\text{.}
\end{equation}
The physical meaning of this condition is that even if the energy
for the incident SP is not resonant with the qubit energy $\hbar
\omega _{0}$, it is still possible to create a highly-entangled
state, but only if the two dots are placed at the right locations.
In this case, the entangled state now becomes
$\xi_{k_{1}}|e_{1},g_{2}\rangle +e^{i\theta }\cdot
\xi_{k_{2}}|g_{1},e_{2}\rangle $, i.e. there is an extra phase
$\theta $ between $|e_{1},g_{2}\rangle $ and $|g_{1},e_{2}\rangle $.

Figure 4 shows the variations of the phase $\theta $ as a function
of the detuning $\delta $. In the limit of large $d$, there is no
super-radiance and the continuous black (top), dashed red, and
dotted blue (bottom) curves represent the results of $\Gamma'=0,$ $
0.025\Gamma _{\textrm{pl}}$, and $0.125\Gamma _{\textrm{pl}}$,
respectively. As can be seen in Fig.~4, once the dissipative loss
$\Gamma'$ is non-zero, the phase suddenly changes from $\pi $ (top
black line) to $0$ (red and blue curves) when $\delta =0$. However,
the scattering treatment we use here does not truly reflect the
dynamics of the coupled system: in reality, it takes a finite time
for the plasmon to mediate the entanglement between the two dots.
During this finite time, the entangled state will undergo
decoherence, even if no plasmon is emitted to the left or right.
Thus this ``high entanglement" value will, in practice, be
reduced (e.g., by emission of the QDs into other modes).

For finite Purcell factor $P$, the entanglement between the two
QDs will be suppressed by the dissipation ($\Gamma'$).  In the
absence of further emission of a surface plasmon (e.g. because of
ideal post-selection, or transfer to the metastable state, see
next section) the entangled state will decay exponentially due to
the losses into free space $\gamma_0$.  This behavior is trivial,
so we do not show it here.  All of this implies the creation of
the entangled state truly depends on $\Gamma_{\textrm{pl}}$ dominating over
the all decay channels (which is feasible~\cite{chang2}), and fast
transfer of the entangled state into a metastable state.  We
discuss this in the next section.

% The black-solid curve in Fig.~5 shows the concurrence for
%$kd=n\pi$ and $\delta=0$ as a function of
%$\Gamma'/\Gamma_{\textrm{pl}}$ $(=1/P)$. As can be seen, the
%concurrence exponentially decreases while increasing
%$\Gamma'/\Gamma_{\textrm{pl}}$. On the other hand, the high
%entanglement ($C\approx1$) is recovered when
%$\Gamma'/\Gamma_{\textrm{pl}}\rightarrow0$ (i.e.
%$P\rightarrow\infty$). We further study the purity to see how the
%dissipation affects the entangled state Eq.~(8). The purity $\wp$
%of the two-dot state can be defined as:
%\begin{equation}
%\wp(\rho)=\textrm{Tr}[\rho^2],
%\end{equation}
%where, $\rho$ is the density matrix of the two-dot state. The
%red-dashed curve in Fig.~5 represents the purity of the two-dot
%state. The purity first decreases rapidly due to the dissipation.
%This means the entangled state is not so ``pure" with finite
%$\Gamma'/\Gamma_{\textrm{pl}}$. However, if
%$\Gamma'/\Gamma_{\textrm{pl}}$ is further increased, the purity
%returns to unity as shown in Fig.~5. This is because most of the
%population decays to the ground state $|g_1,g_2\rangle$ in the
%limit of large $\Gamma'/\Gamma_{\textrm{pl}}$.

%\begin{figure}[th]
%\includegraphics[width=7cm]{fig5.eps}
%\caption{(Color online) The Concurrence of the two-dot
%qubits (black-solid curve) for $kd=n\pi$ and $\delta=0$, and the
%purity (red-dashed curve) of the two-dot state, as functions of
%$\Gamma'/\Gamma_{\textrm{pl}}$ $(=1/P)$. Note that
%$\Gamma'/\Gamma_{\textrm{pl}}=0$ corresponds to the limit of
%infinite Purcell factor $P\rightarrow\infty$.}
%\end{figure}

\section{ENTANGLEMENT STORAGE}
\begin{figure}[th]
\includegraphics[width=\columnwidth]{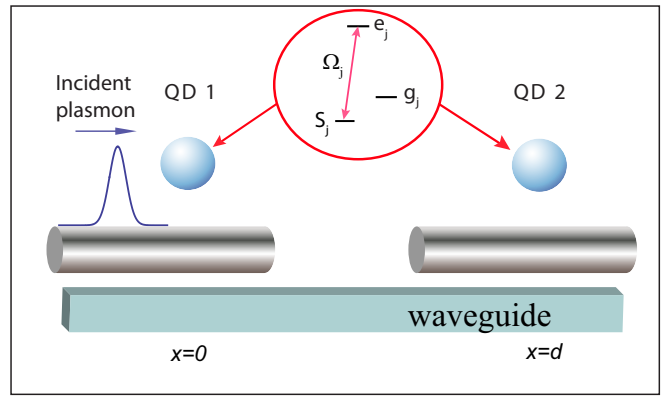}
\caption{(Color online) Schematic diagram of the storage process of the initial
entangled state into metastable entangled states, $\left|
s_{1},g_{2}\right\rangle \pm \left| g_{1},s_{2}\right\rangle$, with
classical optical pulses $\Omega _{1}(t)$ and $\Omega _{2}(t)$. To
avoid the possibility of strong losses in metal nano-wires, a
dielectric waveguide is introduced to achieve remote entanglement.
In the three-level diagram, the subscript $j$ ($j=1$ or $2$) labels
each QD. }
\end{figure}
One might argue that the entangled states created in this manner
are fragile because the QDs are still coupled to the SPs. The
entanglement would rapidly disappear either due to radiative or
non-radiative losses or the SP would eventually escape and be
detected (or lost during propagation in the nano-wire).  The
entangled state exists only on a very short time-scale (see
Appendix B for further discussions on this time-scale). To
overcome this, one can consider multilevel emitters, such as the
three-level configuration\cite{chang2} shown in Fig.~6.
Metastable\cite{Orrit} states, $\left| s_{1}\right\rangle $ and
$\left| s_{2}\right\rangle $, are decoupled from the SPs, but are
resonantly coupled to $\left| e_{1}\right\rangle $ and $\left|
e_{2}\right\rangle $, respectively, via a classical optical
control field with Rabi frequencies $\Omega _{1}(t)$ and $\Omega
_{2}(t)$. Here, the metastable states mean that the relaxation
times of the states $\left| s_{1}\right\rangle$ and
$\left|s_{2}\right\rangle$ are much longer than that of the
excited states $\left| e_{1}\right\rangle$ and
$\left|e_{2}\right\rangle$.

Three remarks about experimental realizations should be addressed
here. First, for the realization of the coupling between a metal
nanowire SP and the two QDs, colloidal CdSe/ZnS QDs and a silver
nanowire are ideal since the exciton energy of CdSe/ZnS QDs is
around 2-2.5 eV, compatible with the saturation plasma
energy\cite{Akimov} ($\approx$ 2.66 eV) of the silver nanowire.

Second, the SPs inevitably experience losses as they propagate
along the nanowire, which limits the feasibility of creating
remote entanglement. One solution to this problem would be to
couple the two dots to two separate nanowires, as shown in Fig.~6.
Also, the wires are evanescently coupled to a phase-matched
dielectric waveguide\cite{chang2,Dayan}. In this case, one can
have both the advantages of strong coupling to the SPs, and
long-distance transport through the dielectric waveguide.

Third, once the entangled state has been prepared, how can it be
detected? One possible procedure is to use ultra-fast optical
tomography as outlined in Ref.~[\onlinecite{Wu}].  This would
allow to reconstruct the density matrix of the effective two-dot
(qubit) system (which in practice will be a mixed state due to
environmental effects, and the inevitable decay of the meta-stable
states), and consequently obtain the concurrence of the two dots.

%Another alternative would be to inject another SP from the left,
%and transfer the entanglement from the quantum dots to the SP
%state. For example, assuming an initial maximally entangled
%(meta-stable) exciton state of
%$\psi_1=\frac{1}{\sqrt{2}}(|g_1,s_2\rangle  + |s_1,g_2\rangle)$,
%applying a selective laser pulse to dot 1 would create the
%short-lived state $\psi_2=\frac{1}{\sqrt{2}}(|e_1,s_2\rangle +
%|s_1,g_2\rangle)$. This state could in principle be transferred to
%an incoming SP which would scatter to give the effective state for
%the SP ``channels" $\psi_p =\frac{1}{\sqrt{2}}(|0_L,1_R\rangle +
%|1_L,0_R\rangle)$. Then, it may be possible to perform a Bell
%inequality violation on this short-lived surface-plasmonic state
%akin to that proposed for edge states in quantum Hall systems.
%However this would require local operations on the $0,1$
%surface-plasmon basis of the left and right channels, perhaps via
%additional quantum dots, or after transferring the SP state to a
%waveguide, and is thus beyond the scope of this work.

\section{SUMMARY AND CONCLUSIONS}
In summary, we have examined the scattering properties of the SPs
in a metal nanowire coupled with two QDs. Both the
dissipative losses and the super-radiant effect are found to
influence the scattering properties.  We then used this system to
propose a scheme to create a remote entangled state between the
two QDs.  Using metastable states and waveguides this might be
possible even in the presence of metal and radiative losses.
 In future work one could consider how the presence of entanglement could
be tested with the Bell Inequality by scattering another plasmon
off the entangled two-dot state and performing scattering
measurements akin to those proposed for edge states in quantum
Hall systems\cite{Beenakker}.

In addition, this proposal can also be applied to other physical
systems. For example, one can easily extend this to photons (e.g.,
in transmission lines or waveguides) coupled to qubits (e.g.,
superconducting qubit)\cite{Shen,zhou,Fan}.

\acknowledgments

This work is supported partially by the National Science Council,
Taiwan, under the grant number NSC 98-2112-M-006-002-MY3. G.Y.C. and
Y.N.C. acknowledge A. Buchleitner, F. Mintert, C. M. Li, and J. R. Johansson for
helpful discussions. N.L. is supported by RIKEN's FPR scheme. F.N.
acknowledges partial support from the Laboratory of Physical
Sciences, National Security Agency, Army Research Office, Defense
Advanced Research Projects Agency, Air Force Office of Scientific
Research, National Science Foundation Grant No. 0726909, JSPS-RFBR
Contract No. 09-02-92114, Grant-in-Aid for Scientific Research (S),
MEXT Kakenhi on Quantum Cybernetics, and Funding Program for
Innovative R\&D on S\&T (FIRST).

\section{Appendix}

\subsection{Super-radiant decay}

%\textbf{From Fermi's Golden Rule, if the inter-dot distance $d$ is
%not large, the spontaneous emission rate $\Gamma$ is proportional
%to the square of the dot-field interaction:
%$\Gamma\propto|\exp(ik_0 r_1)\pm \exp(ik_0 r_2)|^2$, where $r_1$
%$(r_2)$ is the position of dot-1 (2). Since we have set $r_1=0$
%and $|r_2-r_1|\equiv d$, we can further write
%$\Gamma\propto\bigg[1\pm \frac{\sin(k_{0}d)}{k_{0}d}\bigg]$.
%Therefore, in the limit of $d\rightarrow0$, $\Gamma$ can be
%enhanced to twice larger (super-radiance) or suppressed to zero
%(sub-radiance). The second line of Eq.~(12) represents the effect
%of the collective decay (super-radiance)\cite{ynchen}}.

The inclusion of the super-radiant term in the non-Hermitian form
can be justified as follows. One can start with a master equation
for the two QDs coupled to external modes or with a simple
application of Fermi's Golden rule (e.g., as in Ref.
[\onlinecite{devoe}]). In the singlet, $\left| S\right\rangle
=\frac{1}{\sqrt{2}}(\left| e_{1},g_{2}\right\rangle -\left|
g_{1},e_{2}\right\rangle)$ , and triplet, $\left| T\right\rangle
=\frac{1}{\sqrt{2}}(\left| e_{1},g_{2}\right\rangle +\left|
g_{1},e_{2}\right\rangle $, bases this master equation has several
super-radiant and sub-radiant decay terms, describing the
following decay channels:  via rate $\Gamma_+$ for the channel
$T_+ \rightarrow T \rightarrow T_-$; and via the rate $\Gamma_-$
for the channel $T_+ \rightarrow S \rightarrow T_-$. Since we omit
the double-occupation terms here, only the channels from $S$ and
$T$ contribute. These give the following Lindblad terms in the
master equation for the density matrix of the two
dots\cite{ynchen,devoe},
\begin{eqnarray}
L_+&=&-\frac{\Gamma_+}{2}\bigg(
|T\rangle\langle T|\rho+\rho|T\rangle\langle T|-|T_-\rangle\langle T|\rho|T\rangle\langle T_-|\bigg), \notag\\
L_-&=&-\frac{\Gamma_-}{2} \bigg(|S\rangle\langle
S|\rho+\rho|S\rangle\langle S|-|T_-\rangle\langle
S|\rho|S\rangle\langle T_-|\bigg).\notag\\
\end{eqnarray}
Where
\begin{equation}
\Gamma_{\pm} = \gamma_0\bigg[1\pm \frac{\sin(k_{0}d)}{k_{0}d}\bigg].
\end{equation}
For both $L_-$ and $L_+$, the first two terms contribute to the
non-hermitian Hamiltonian, while the last term is the ``quantum
jump" term.  As mentioned above, these jump terms can be neglected
in here because a loss of excitation from either quantum dot means a
loss of the input SP. Combining the other terms, and rewriting the
singlet and triplet bases in the dot excitation basis, gives the
following contribution to the effective non-Hermitian Hamiltonian,
\begin{eqnarray}
H_{\textrm{eff}}^{\textrm{SR}}&=&-i \hbar \frac{\gamma_0}{2}
\left(\sigma_{e_1,e_1} + \sigma_{e_2, e_2}\right)\notag\\
&-&i \hbar \frac{\sin(k_{0}d)}{2k_{0}d}\gamma _{0}\left(\sigma^{(1)}_+\sigma^{(2)}_-+\sigma^{(1)}_-\sigma^{(2)}_+\right)
\end{eqnarray}

\subsection{ENTANGLEMENT STORAGE} For the entanglement
storage process, instead of transforming Eq.~(\ref{H1}) into real
space, we represent the Hamiltonian under the bases of singlet,
$\left| S\right\rangle =\frac{1}{ \sqrt{2}}(\left|
e_{1},g_{2}\right\rangle -\left| g_{1},e_{2}\right\rangle )$ , and
triplet, $\left| T\right\rangle =\frac{1}{\sqrt{2}}(\left|
e_{1},g_{2}\right\rangle +\left| g_{1},e_{2}\right\rangle $,
states:

\begin{eqnarray}
H &=&\hbar \left(\omega _{0}-i\frac{\Gamma ^{\prime
}}{2}\right)\left(\left| T\right\rangle \left\langle T\right|
+\left| S\right\rangle \left\langle S\right| \right)  \notag
\\
&&-\hbar g\int
dk~\left\{\left[\frac{1}{\sqrt{2}}\left(1+e^{ikd}\right)\left|
T\right\rangle
\left\langle g_{1},g_{2}\right| a_{k} \right.\right. \notag \\
&&\left.\left.+\frac{1}{\sqrt{2}}\left(1-e^{ikd}\right)\left|
S\right\rangle \left\langle
g_{1},g_{2}\right| a_{k}\right]+\textrm{h.c.}\right\}  \notag \\
&&+\int dk~\hbar v_{\textrm{g}}|k|a_{k}^{\dag }a_{k},\label{h3}
\end{eqnarray}
where we have adopted the approximation that the super-radiant
effect can be neglected in the limit of large $d$. We now consider
the general time-dependent wave function
\begin{eqnarray}
\left| \psi \right\rangle  &=&\int
dk[c_{R,k}(t)\hat{a}_{R,k}^{\dagger
}+c_{L,-k}(t)\hat{a}_{L,-k}^{\dagger }]\left|
g_{1},g_{2};0\right\rangle   \notag \\
&&+c_{T}(t)\left| T;0\right\rangle +c_{S}(t)\left| S;0\right\rangle  \label{psi2}\\
&&+c_{M_{T}}(t)\left| M_{T};0\right\rangle +c_{M_{S}}(t)\left|
M_{S};0\right\rangle ,  \notag
\end{eqnarray}
where $\left| M_{S}\right\rangle =\frac{1}{\sqrt{2}}(\left|
s_{1},g_{2}\right\rangle -\left| g_{1},s_{2}\right\rangle )$ and,
$\left| M_{T}\right\rangle =\frac{1}{\sqrt{2}}(\left|
s_{1},g_{2}\right\rangle +\left| g_{1},s_{2}\right\rangle )$, denote
the singlet and triplet metastable states, respectively. From $H$
$\left| \psi \right\rangle =i\hbar \partial_t\left| \psi
\right\rangle $, the state amplitudes evolve according to

\begin{eqnarray}
\frac{d~c(t)}{dt} &=&-i\delta _{k}c(t)+\frac{ig
}{\sqrt{2}}(1+e^{-ikd})c_{T}(t)  \notag \\
&&+\frac{ig}{\sqrt{2}}(1-e^{-ikd})c_{S}(t),
\end{eqnarray}
where $\delta _{k}=v_\textrm{g}k-\omega _{0}$. The equation above
refers to, in a compact manner, to two equations: one for
$c_{R,k}$ and the other one for $c_{L,-k}$, respectively. If
$\Omega _{1}(t)=\Omega _{2}(t)$ and $ kd=2n\pi $, Eq. (\ref{h3})
can be substituted into the equation of motion for $c_{T}(t)$

\begin{eqnarray}
\frac{d~c_{T}(t)}{dt} &=&-\frac{\Gamma ^{\prime
}}{2}c_{T}(t)+i\Omega
_{1}(t)c_{M_{T}}(t)  \notag \\
&&+ig\int dk[c_{R,k}(t)+c_{L,-k}(t)],
\end{eqnarray}
which yields an integro-differential equation involving $c_{T}(t)$.
Imposing the reasonable constraint that the SP storage process has
no outgoing field at the end of the wire, such that
$c_{R,k(L,-k)}(\infty )=0$, one can obtain an implicit expression
for the required pulse shape $\Omega _{1}(t)$ and the following
equation related to the population in the state $\left|
M_{T}\right\rangle$

\begin{equation}
\frac{d}{dt}\left| c_{M_{T}}(t)\right|
^{2}=\frac{-v_{\textrm{g}}^{2}}{2\pi g^{2}}\bigg(\frac{d}{ dt}\left|
E_{T}(t)\right| ^{2}-\frac{\Gamma _{pl}-\Gamma ^{\prime }}{2} \left|
E_{T}(t)\right| ^{2}\bigg),
\end{equation}
where
\begin{equation}
E_{T}(t)=-\sqrt{2\pi }igc_{T}(t)/v_{\textrm{g}}.
\end{equation}
With the normalization condition,
\begin{equation}
\int_{-\infty }^{\infty }dt\left| E_{T}(t)\right|
^{2}=1/(2v_{\textrm{g}}),
\end{equation}
and assuming that the incoming field vanishes\cite{chang2} at
$t=\pm \infty $ [i.e., $ E_{T}(\pm \infty )=0$], Eq.~(\ref{psi2})
can be integrated to yield
\begin{equation}
\left| c_{M_{T}}(\pm \infty )\right| ^{2}=1-1/P,
\end{equation}
where
\begin{equation}
P\equiv\Gamma _{\textrm{pl}}/ \Gamma ^{\prime },
\end{equation}
is the effective Purcell factor. Similarly, it can be easily shown
that the storage efficiency into the $\left| M_{S}\right\rangle $
state is also equal to $1-1/P$, if $\Omega _{1}(t)=-\Omega _{2}(t)$
and $ kd=(2n+1)\pi $. Note that the metal and radiative losses on
the qubits are taken into account in the above derivation.
Therefore, the entangled states can be stored with a high efficiency
only if the Purcell factor $P$ is large enough. Furthermore, the two
qubits could be separated far away from each other, such that one
can address one qubit without affecting the other.

\end{document}